\documentclass[twocolumn,showpacs,amsmath,amssymb,prx,aps]{revtex4}   %%  4-pages !!
\usepackage{graphicx}% Include figure files
\usepackage{dcolumn}% Align table columns on decimal point
\usepackage{bm}% bold math
\usepackage{color}% color text
\usepackage{amsmath}

\renewcommand{\vec}[1]{\mathbf{#1}}

\newif\ifgraph

\graphtrue             %
\begin{document}
\title{Active Particle Diffusion in Convection Roll Arrays}

\author{Pulak K. Ghosh$^{1}$}
\author{Fabio Marchesoni$^{2,3}$}
\author{Yunyun Li$^{2}$}
\email{yunyunli@tongji.edu.cn}
\author{Franco Nori$^{4,5}$}

\affiliation{$^{1}$ Department of Chemistry, Presidency University, Kolkata
700073, India}
 \affiliation{$^{2}$ Center for Phononics and Thermal Energy Science, Shanghai
 Key Laboratory of Special Artificial Microstructure Materials and Technology,
 School of Physics Science and Engineering, Tongji University, Shanghai 200092, China}
 \affiliation{$^{3}$ Dipartimento di Fisica, Universit\`{a} di Camerino, I-62032 Camerino, Italy}
 \affiliation{$^{4}$ Theoretical Quantum Physics Laboratory,
RIKEN Cluster for Pioneering Research, Wakoshi, Saitama 351-0198, Japan}
\affiliation {$^{5}$ Department of Physics, University of Michigan, Ann
Arbor, Michigan 48109, USA}

\date{\today}

\begin{abstract}
We numerically investigated the Brownian motion of active Janus particles
in a linear array of planar counter-rotating convection rolls at high
P\'eclet numbers. Similarly to passive particles, active microswimmers
exhibit advection enhanced diffusion, but only for self-propulsion speeds
up to a critical value. The diffusion of faster Janus particles is governed
by advection along the array's edges, whereby distinct diffusion regimes
are observed and characterized. Contrary to passive particles, the relevant
spatial distributions of active Janus particles are inhomogeneous. These
peculiar properties of active matter are related to the combined action of
noise and self-propulsion in a confined geometry and hold regardless of the
actual flow boundary conditions.

\end{abstract}
\maketitle

\section {Introduction} \label{Intro}

The diffusion of a tracer (organic or artificial, alike) in a suspension
fluid is a standard problem of classical transport theory \cite{Redner}. This
paper combines two distinct aspects of this phenomenon, which recently
attracted widespread interdisciplinary interest, each for its own merit: (i)
the persistent (or time-correlated) random motion of self-propelling
particles and (ii) colloidal dispersion in laminar flows.

The most tractable example of persistent Brownian motion is represented by
artificial micro-swimmers, namely tiny Brownian particles capable of
self-propulsion in an active medium \cite{Granick,Muller}.
%Such particles are designed to harvest environmental energy by converting
%it into kinetic energy.
A class of artificial swimmers widely investigated in the current literature
is the so-called Janus particles (JP), mostly spherical colloidal particles
with two differently coated hemispheres, or ``faces''
\cite{Marchetti,Gompper}. Recently, artificial micro- and nano-swimmers of
this class have been the focus of pharmaceutical (e.g., smart drug delivery
\cite{smart}) and medical research (e.g., robotic microsurgery \cite{Wang}).
These peculiar Brownian particles change direction randomly as usual, but
with finite time scale; persistence makes their diffusion extremely sensitive
to geometric confinement and other constraints \cite{ourPRL,sperm}.
Technological applications involving sub-millimeter artificial swimmers thus
require accurate control of their diffusive properties in non-homogeneous
environments \cite{Redner,BechRMP}.

\begin{figure*}[tp]
\centering \includegraphics[width=7.7cm]{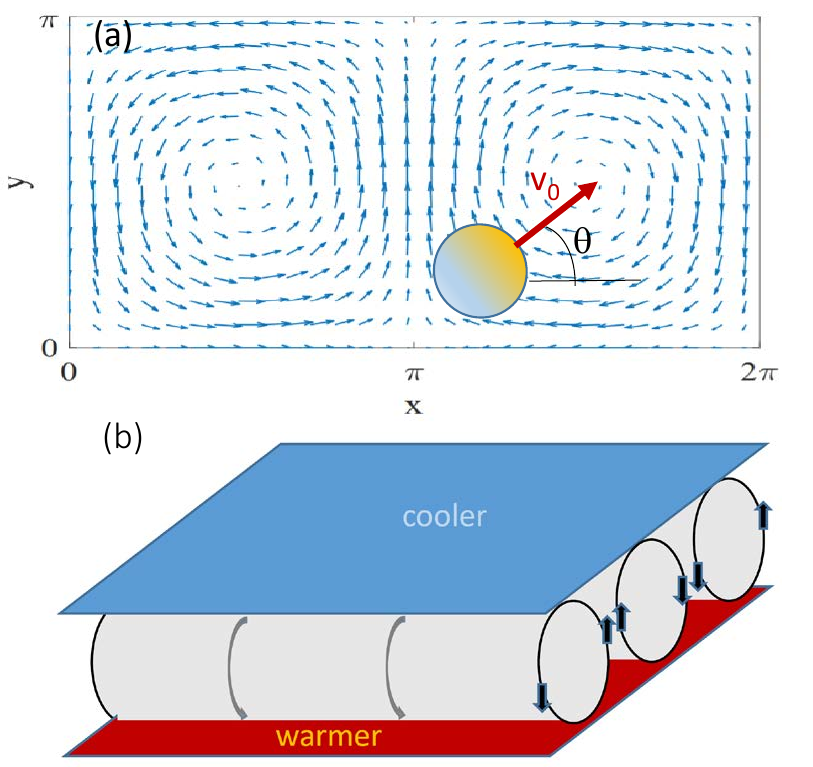}
\centering \includegraphics[width=8.1cm]{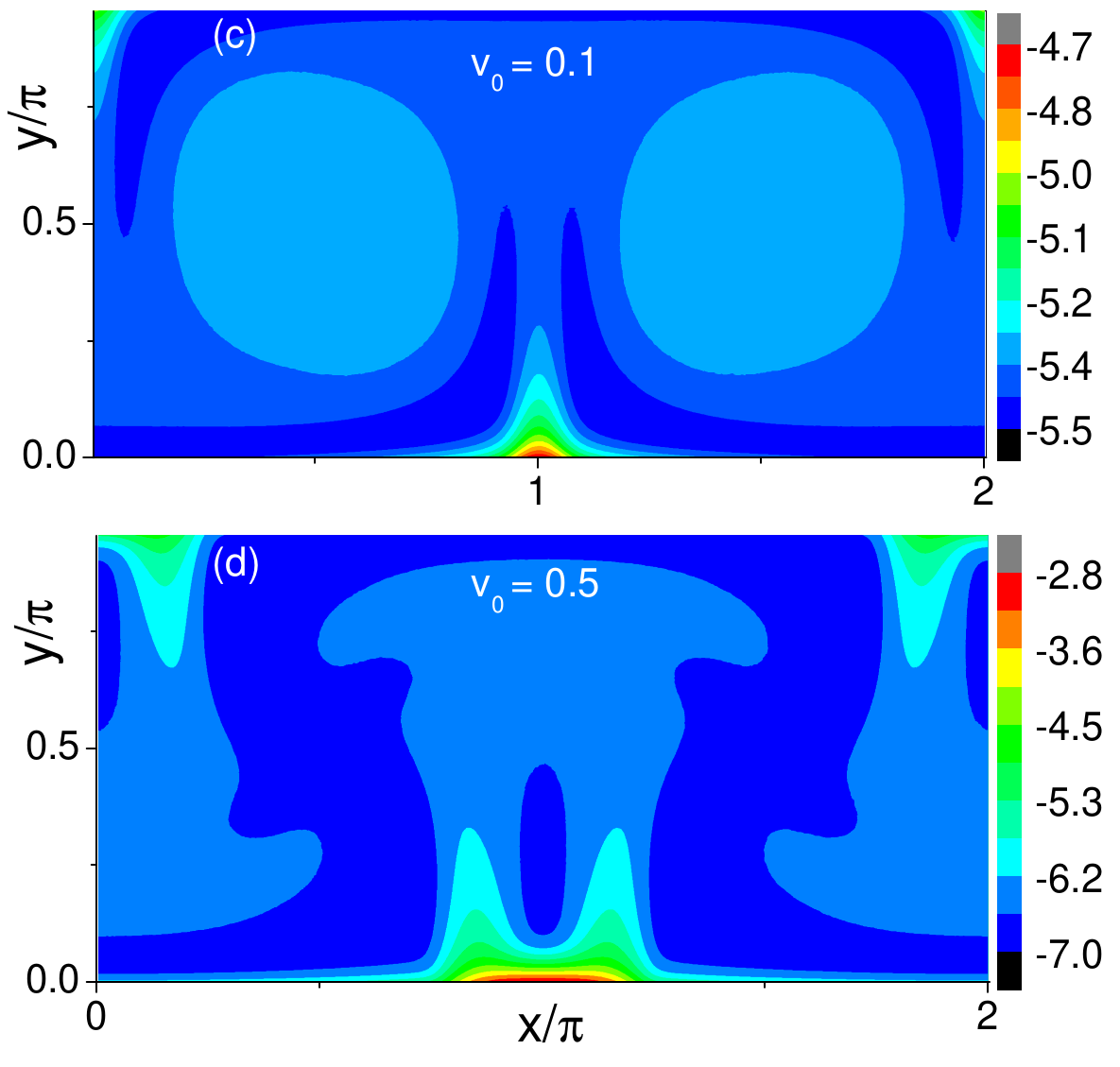}
\caption{Spatial distributions of a Janus particle in the laminar flow of Eq.~(\ref{psif}), sketched in (a)-(b), for
$v_0=0.1$ (c) and $0.5$ (d). The chart levels are color-coded on natural logarithmic
scales as indicated. Other simulation parameters are: $D_0=0.01$, $D_\theta=0.01$,
$U_0=1$ and $L=2 \pi$. According to Eq.~(\ref{vc}),
here $v_c=0.4$. A practical realization of a linear convection array is represented by
the Rayleigh-B\'enard rolls sketched in (b); the JP self-propulsion model
of Eq.~(\ref{LE}) is illustrated in (a).}
\label{F1}
\end{figure*}

On the other hand, Brownian diffusion in an advective medium is also a
nanotechnological issue, for instance, in the design and operation of
microfuidic devices \cite{Kirby,Tabeling,PNAS} or chemical reactors
\cite{Moffatt}. Let us consider a Brownian tracer of free diffusion constant
$D_0$, advected by the free-boundary stationary laminar flow of Fig.~\ref{F1}(a) with stream function \cite{Chandra,Child2},
\begin{equation}
\label{psif}
\psi(x,y)= ({U_0L}/{2\pi})\sin({2\pi x}/{L})\sin({2\pi y}/{L}).
\end{equation}
On combining the two constants, $L$, the flow's spatial period, and $U_0$,
the maximum advection speed, one defines the advection diffusion scale,
$D_L=U_0L/2\pi$, and the maximum roll vorticity, $\Omega_L=2\pi U_0/L$
(Appendix A).

At high P\'eclet numbers, ${\rm Pe}=D_L/D_0\gg 1$, a passive tracer undergoes
normal diffusion with enhanced diffusion constant  $D=\kappa \sqrt{D_LD_0}$
with $\kappa=1.065$, that is $D>D_0$ \cite{Rosen}. This advection effect,
termed advection enhanced diffusivity (AED), has been explained
\cite{Rosen,Soward,Shraiman,Pomeau} by noticing that for $D_0<D_L$ an
unbiased particle jumps between convection rolls while being advected along
their separatrices. Narrow flow boundary layers (FBL) of estimated width
$\delta =({D_0/\Omega_L})^{1/2}$, form a network of advection channels
centered around the $\psi(x,y)$ cell separatrices, thus enabling a
large-scale particle's diffusion.

Peculiar effects due to the combination of self-propulsion and advection are
expected to emerge when one considers an active JP suspended in a one
dimensional (1D) array of counter-rotating convection rolls. An ideal
experimental setup is sketched in Fig.~\ref{F1}(b). An array of stationary
Rayleigh-B\'enard cells can occur in a plane horizontal layer of fluid heated
from below \cite{RB1,RB2}. Assuming that they are counter-rotating cylinders
parallel to the $z$-axis, the $z$ coordinate of a suspended tracer is
ignorable; hence the reduced two dimensional (2D) flow pattern of Eq.
(\ref{psif}). Advection enhanced diffusivity of passive colloidal particles in arrays of
Rayleigh-B\'enard rolls has already been demonstrated experimentally
\cite{Gollub1,Gollub3}. Experimental data on the dispersion of
self-propelling microswimmers in convective laminar flows are scarse. In this
regard, active JPs are ideal tracers for this kind of measurements because
self-propulsion speed can be conveniently tuned with respect to the advection
drag established in the convection cell.

This paper is organized as follows. In Sec. \ref{Model} we present our model
and briefly discuss the dynamical significance of the relevant parameters.
Our derivation of the relevant time scales is detailed in Appendix A. Our
main numerical results are analysed in Secs. \ref{px} and \ref{Dx}, where we
show that: (i) The interplay of advection and self-propulsion causes the
nonuniform spatial distribution of a confined active JP. For self-propulsion
speeds below a certain threshold, its distribution tends to accumulate along
the roll boundaries (Sec. \ref{px} and Appendix B); (ii) Under these
conditions, its self-propulsion and advection velocities tend to line up, so
that, contrary to the 2D case of Ref. \cite{RR2}, the large-scale diffusion
of an active JP is insensitive to self-propulsion itself (Sec. \ref{Dx});
(iii) Active tracers with self-propulsion speeds larger than the above
threshold, attain a maximum diffusion constant for an optimal persistence
time, which we relate to advection at the array's edges (Sec. \ref{Dx} and
Appendix C). In Sec. \ref{Conclusions} we stress the role of geometric
confinement on the diffusion properties of an active JP in a convection array
and show that the picture above holds also for rigid (i.e., no-slip) edge
flows.

\section{Model} \label{Model}

By (linear) convection array we mean here a stationary laminar flow with
periodic stream function like $\psi(x,y)$ of Eq.~(\ref{psif}), confined
between two parallel edges, $y=0$ and $y=L/2$, which act as dynamical
reflecting boundaries. The unit cell of the array consists of two
counter-rotating convection rolls [Fig.~\ref{F1}(a)]. The dynamics of an
overdamped active JP can then be formulated by means of two translational and
one rotational Langevin equation (LE),
\begin{eqnarray} \label{LE}
\dot {\vec r}&=& {\vec v}_\psi + {\vec v}_0 +\sqrt{D_0}~{\bm \xi}(t) \\ \nonumber
\dot \theta &=& ({\alpha}/{2})~\nabla \times {\vec v}_\psi +\sqrt{D_\theta}~\xi_\theta (t),
\end{eqnarray}
where ${\vec r}=(x,y,)$, ${\vec v}_\psi =(\partial_y, -\partial_x)\psi$
denotes the advection velocity and the self-propulsion vector, ${\vec
v}_0=v_0(\cos \theta, \sin \theta)$, has constant modulus, $v_0$, and is
oriented at an angle $\theta$ with respect to the longitudinal $x$-axis.
The translational (thermal) noises in the $x$ and $y$ directions, ${\bm
\xi}(t)=(\xi_x(t), \xi_y(t))$, and the rotational noise, $\xi_\theta (t)$,
are stationary, independent, delta-correlated Gaussian noises, $\langle
\xi_i(t)\xi_j(0)\rangle = 2 \delta_{ij}\delta (t)$, with $i,j=x,y,\theta$.
$D_0$ and $D_\theta$ are the respective noise strengths, which for generality we assume to
be unrelated~\cite{ourPRL}. To avoid uncontrolled hydrodynamic
effects, the particle is taken to be pointlike \cite{PNAS}. Other effects due
to its actual geometry and chemical-physical characteristics are encoded in
the model dynamical parameters. The reciprocal of $D_\theta$ coincides with
the angular persistence (or correlation) time, $\tau_\theta$, of  ${\vec
v}_0$; accordingly, $l_\theta=v_0/D_\theta$ quantifies the persistence length
of the particle's self-propelled random motion. The flow shear exerts a
torque on the particle proportional to the local fluid vorticity, $\nabla
\times {\vec v}_\psi$ \cite{Neufeld,RR2}. For simplicity, we adopt Fax\'en's
second law, which, for an ideal no-stick spherical particle, yields
$\alpha=1$ \cite{Stark}.  In the high P\'eclet number regime addressed here,
${\rm Pe} \gg 1$ or $D_0 \ll D_L$, particle diffusion is strongly influenced
by advection (appendix A).

\begin{figure}[tp]
\centering \includegraphics[width=8.0cm]{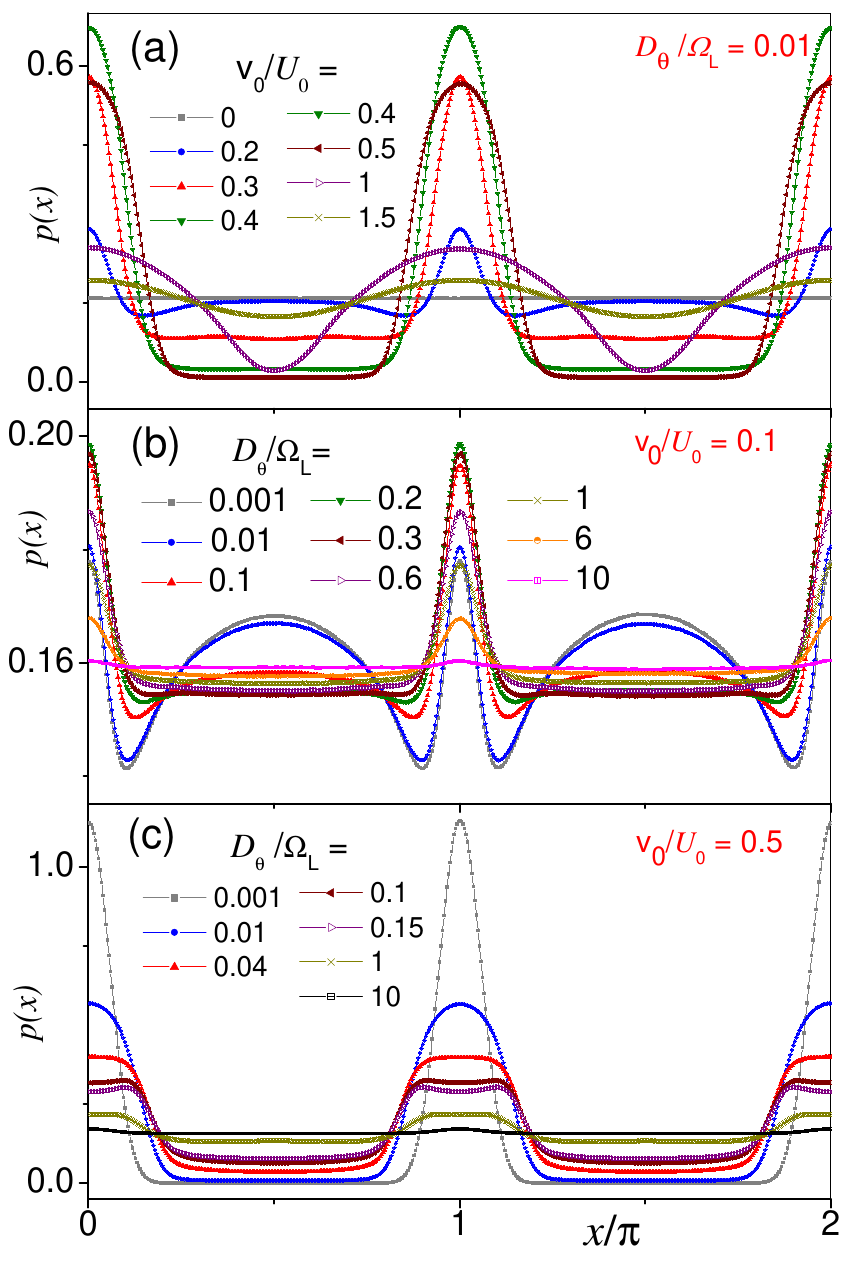}
\caption{Stationary longitudinal distributions, $p(x)$, of a Janus particle in the laminar flow of Eq.~(\ref{psif}) for
(a) $D_\theta=0.01$ and different $v_0$; (b) $v_0=0.1$ and (c) $v_0=0.5$
and different $D_\theta$ (see legends).
Other simulation parameters are: $D_0=0.01$; $U_0=1$ and $L=2 \pi$, with $D_L=\Omega_L=1$.}
\label{F2}
\end{figure}

The Langevin equation (\ref{LE}) can be conveniently reformulated in dimensionless units by
rescaling $(x,y) \to (\tilde x, \tilde y)=(2\pi/L)(x,y)$ and $t \to \tilde t=
\Omega_L t$. The three remaining independent parameters get rescaled as
follows: $v_0 \to v_0/U_0$, $D_0 \to D_0/D_L$ and $D_\theta \to
D_\theta/\Omega_L$. This means that, without loss of generality, we can set
$L=2\pi$ and $U_0=1$ and the ensuing simulation results can be regarded as
expressed in dimensionless units and easily scaled back to arbitrary
dimensional units. The stochastic differential Eqs.~(\ref{LE}) were
numerically integrated by means of a standard Milstein scheme \cite{Kloeden}.
Particular caution was exerted when computing  the asymptotic diffusion
constant, $D=\lim_{t\to \infty} \langle [x(t) -x(0)]^2\rangle /2t,$ because
for low values of the noise strengths, $D_0$ and $D_\theta$, the transients
of the diffusion process grow exceedingly long \cite{Neufeld,ourPOF}. For
asymptotically large running times, our estimates of $D$ are independent of
the starting point $(x(0),y(0))$.

\section{Spatial distributions} \label{px}

In sharp contrast with the noiseless limit, $D_0=D_\theta =0$, investigated
in Ref. {\cite {Neufeld}}, the spatial distribution of a noisy active JP is
not uniform. The outcome of our numerical simulations is summarized in Figs.
\ref{F1}(c),(d) and \ref{F2}. The laminar flow acts upon the particle through
both an advection drag and an advection torque. Along the roll boundaries the
drag is maximum (with speed approaching $U_0$, except at the ``stagnation''
corners), but the torque vanishes. At low self-propulsion speeds, this
favours the orientation of $\vec{v}_0$ parallel to the advection velocity
$\vec{v}_\psi$. The JP thus undergoes a large-scale intra-roll circulation
motion, which causes its accumulation along the outer layers of the rolls.
The less pronounced particle accumulation at the roll centers is attributable
to the higher vorticity there \cite{ourPOF}. These two areas of accumulation
are separated by a circular depletion region. Indeed, in Figs. \ref{F1}(c)
and Figs. \ref{F2}(a),(b) (see also Appendix B) the particle appears to be
sucked in by the ascending ($x=L/2$) and descending boundary flows ($x=0,
L/2$), an effect that seems to increase with increasing $v_0$.

This picture changes abruptly as $v_0$ is raised above a critical value $v_c$
[Fig.~\ref{F2}(a)], which we established to depend on the strength of the
thermal noise, $D_0$ (Appendix C). The intra-roll circulation of Fig.~\ref{F1}(c)
 is suppressed and the roll interior gets depleted [Fig.~\ref{F2}(a),(c)]; as a result,
  the particle piles up symmetrically at the
base of the ascending (bottom edge) and descending flows (top edges).
Moreover, for $v_0\gtrsim U_0$, the particle seems to diffuse mostly along
the array's edges, which explains why the longitudinal distributions, $p(x)$,
turn uniform again with increasing $v_0$, while the transverse distributions,
$p(y)$, remain peaked at $y=0, L/2$ (Appendix B). One also notices that the
peaks of $p(x)$ widen with increasing $v_0$ [Fig.~\ref{F2}(a)] and $D_\theta$
[Fig.~\ref{F2}(c)].

The relevance of these results can be best appreciated by comparison with the
diffusion of a passive particle in the same 1D convection array. In that
case, the flow boundary layers still control the particle's large-scale diffusion, but all
stationary distributions, $p(x,y)$, remain uniform \cite{ourPOF}. This
conclusion applies also to noiseless self-propelling JPs in 1D convection
arrays, as proven in Ref. \cite{Neufeld}, but is no longer true in the
presence of thermal noise. Indeed, upon hitting either array edge, the
particle will persist pointing against it for a time $\tau_\theta$; hence the
angular correlation of ${\vec v}_0$ and ${\vec v}_\psi$.  [Note that in most
simulations presented here $\tau_\theta$ is larger than the circulation
characteristic time, ie, $D_\theta <\Omega_L$.] Accordingly, no probability
density accumulation at the roll boundaries was detected for an active JP in
the 2D laminar flow of Eq.~(\ref{psif}) with periodic boundary conditions,
regardless of the noise strengths, $D_0$ and $D_\theta$ (Appendix B). This
leads to the conclusion that the FPL structure we detected in the stationary
distributions $p(x,y)$ of an active JP diffusing in a 1D convection array is
{\em a combined effect of noise and geometric confinement}.

\begin{figure}[tp]
\centering \includegraphics[width=8.0cm]{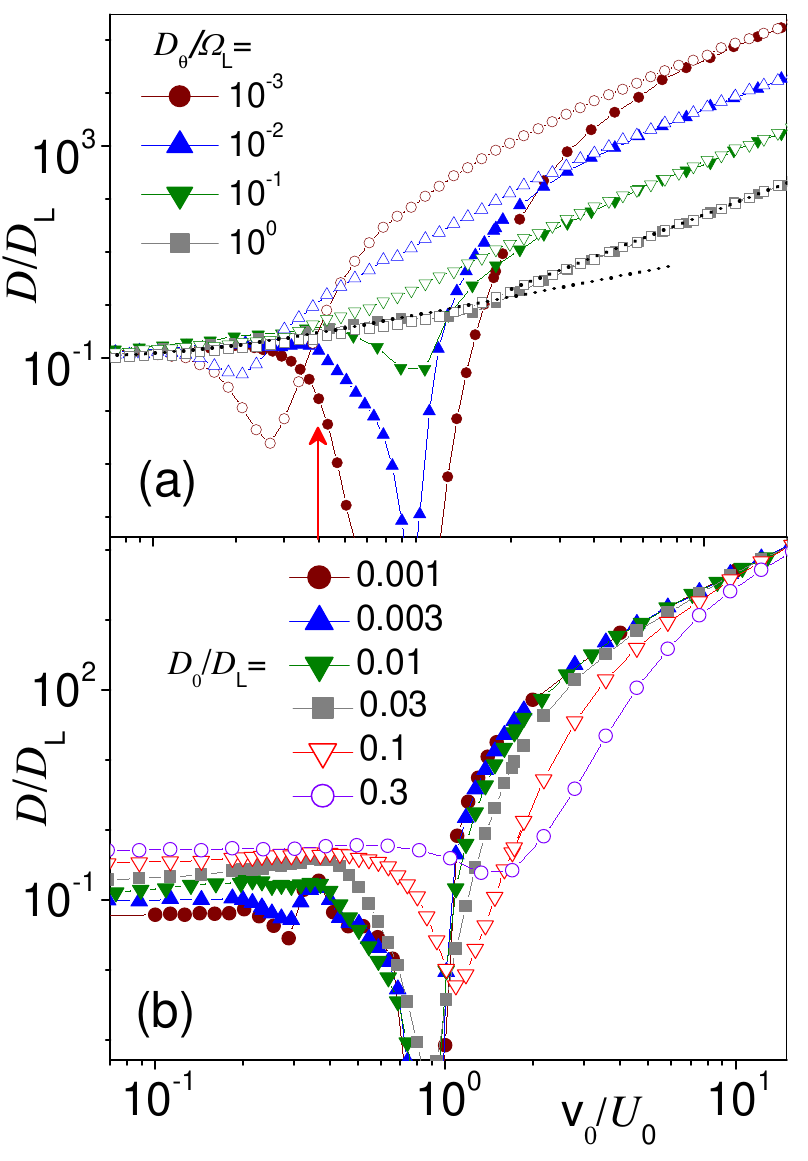}
\caption{(a) Longitudinal diffusion of a JP in the laminar flow of Eq.~(\ref{psif}) (solid symbols)
and (\ref{psir}) (empty symbols): $D/D_L$
vs. $v_0/U_0$ for $D_0=0.01$ and different $D_\theta$ (see legends).
Dashed curves represent the estimates,
$D=\kappa \sqrt{D_L (D_0+D_s)}$ and
$D=D_0+D_s$, with $D_s=v_0^2/2 D_\theta$, respectively, for low and high $v_0$ (see text).
Our estimate for $v_c$, Eq.~(\ref{vc}), is marked by a vertical arrow.
(b) $D/D_L$ vs. $v_0/U_0$ for $D_\theta=0.01$ and different $D_0$. Flow
parameters are $U_0=1$ and $L=2 \pi$, with $D_L=\Omega_L=1$.} \label{F3}
\end{figure}

The self-propulsion threshold, $v_c$, can be estimated as follows. When the
vector $\vec{v}_0$ points inwards, the particle pulls away from the edge a
length of the order of $v_0/4\Omega_L$, before being swept into a vertical
flow layer. As such length grows comparable with the width of an unbiased
flow boundary layers, i.e., for $v_0> v_c$ with
\begin{equation} \label{vc}
v_c/U_0=4 \sqrt{D_0/D_L},
\end{equation}
the particle exits the FBL and its circulation along the roll separatrices is
interrupted. This estimate of $v_c$ is consistent with our simulation data
for $p(x)$ and $p(y)$ at low angular noise, $D_\theta \ll \Omega_L$ [compare
Figs. \ref{F1} and \ref{F2}; see Appendices B and C for more details]. Note,
for instance, that in Fig.~\ref{F2} the $p(x)$ regions delimited by the peaks
at $x=0, \pi$ and $2\pi$ get depleted only for $v_0=0.5$, that is for
$v_0>v_c$. Moreover, being confined in a FBL, a JP with $v_0<v_c$ ought to
behave like a passive colloidal particle, ie, undergo advection enhanced 
diffusivity as an effect of the
sole thermal noise. The diffusion data presented in the next section (Fig.~
\ref{F3}) confirm this conclusion.

As the FBL circulation breaks up, the JP tends to accumulate against the
array edges, provided that the self-propulsion length is larger than the
array width, $l_\theta >L/2$, or, equivalently, $D_\theta/\Omega_L <
v_0/U_0$. However, its motion along the edges is not advection-free. The
coordinate $x$ in Eq.~(\ref{LE}) then obeys the approximate LE, $\dot x=U_0
\langle \cos(2\pi y/L)\rangle \sin (2\pi x/L) +v_0 \cos \theta +\xi_x(t)$,
which describes the dynamics of a Brownian particle pinned to a washboard
potential \cite{Risken} (advection term) and subjected to a colored,
non-Gaussian tilting noise, $v_0 \cos \theta(t)$, with correlation time
$\tau_\theta$ \cite{ourPRL} (self-propulsion term). The average $\langle
\cos(2\pi y/L)\rangle$ depends on all three free parameters $v_0, D_\theta$
and $D_0$; in particular, its modulus increases with increasing $v_0$ and
decreasing $D_\theta$. This simple observation explains: (i) the
non-monotonic $v_0$-dependence of the $p(x)$ peaks, whereby a larger $v_0$
implies not only higher washboard potential barriers, but also a stronger
tilting term; (ii) the flattening of the longitudinal distributions for $v_0
\gtrsim U_0$, as self-propulsion wins over the advection pinning action at
the edges; (iii) the broadening and double-peaked profile of the $p(x)$ peaks
in Fig.~\ref{F2}(c) on increasing $D_\theta$ which is a well-known effect of
colored noise \cite{color}.

On increasing $D_\theta$, the JP self-propulsion length eventually grows
shorter than the roll size, $l_\theta<L/2$; the active particle then tends to
behave like a passive Brownian particle, except its free diffusion constant,
$D_0$, must be now incremented by the extra term $D_s=v_0^2/2D_\theta$.
Accordingly, both its spatial distributions, $p(x)$ and $p(y)$, become
uniform [see Figs. \ref{F2}(b),(c) and Appendix B].

\begin{figure}[tp]
\centering \includegraphics[width=8.0cm]{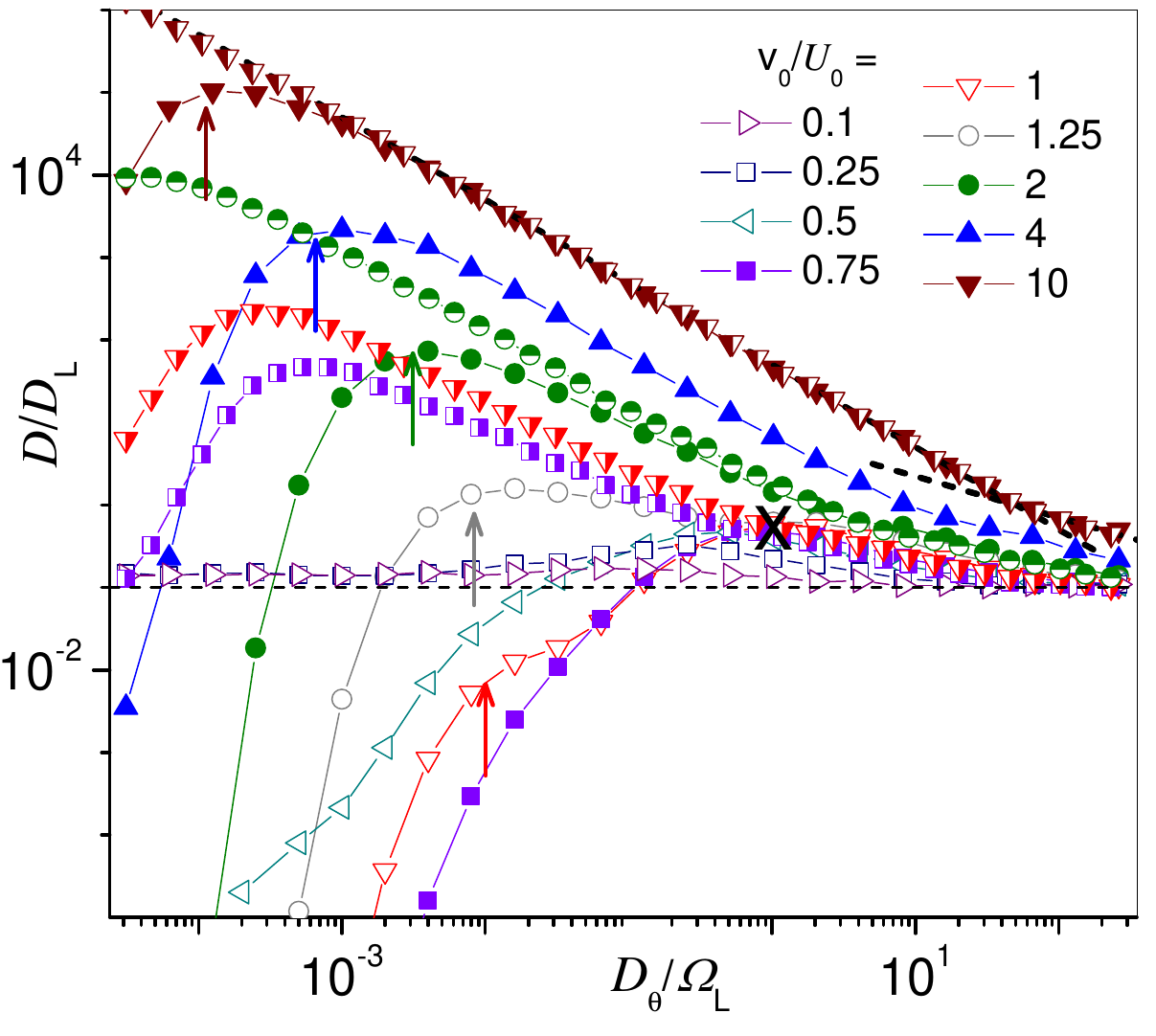}
\caption{Longitudinal diffusion of a JP in the laminar flow of Eq.~(\ref{psif}): $D/D_L$
vs. $D_\theta/\Omega_L$ for different values of $v_0$ (filled and empty symbols, see legends).
We remind that here Eq.~(\ref{vc}) yields $v_c=0.4$.
Dashed curves represent the limiting cases
$D=\kappa \sqrt{D_L D_0}$ (passive particle with ${\rm Pe} \gg 1$), $D=D_0+D_s$ (free JP)
and  $D=\kappa \sqrt{D_L (D_0+D_s)}$ (advection enhanced diffusivity of a JP in 2D \cite{RR2}),
 with $D_s=v_0^2/2 D_\theta$ (see text).
Vertical arrows mark our estimates for $D_\theta ^*$, Eq.~(\ref{Db}), and a cross the maxima,
$D/D_L=0.5$, of the curves with $v_c < v_0 \lesssim U_0$.
Other simulation parameters are: $D_0=0.01$, $U_0=1$ and $L=2 \pi$, hence $D_L=\Omega_L=1$.
A few curves for the same JP in the rigid-boundary flow of Eq.~(\ref{psir})
are plotted for a comparison (half-filled symbols of the corresponding shape and color).}
\label{F4}
\end{figure}

\section{Longitudinal diffusion} \label{Dx}

Based on the qualitative arguments of Sec. \ref{px}, we expect to observe
distinct diffusion regimes for a JP with $l_\theta > L/2$. Our expectation
are supported by the simulation data reported in Fig.~\ref{F3}(a,b). Indeed, the
curves $D$ versus $v_0$ exhibit distinct behaviors for $v_0<v_c$, $v_c < v_0
\lesssim U_0$ and $v_0 \gg U_0$. For $v_0 \gg U_0$, advection is negligible
compared to self-propulsion; since we assumed reflecting boundaries at the
array's edges, not surprisingly, $D\to D_0+D_s$ \cite{RR2}. This behavior is
in sharp contrast with the scenario suggested by the $D$ curves in the limit
$v_0/U_0 \to 0$. All curves overlap, insensitive to $D_\theta$, and, more
remarkably, tend to the advection enhanced diffusivity estimate, $D=\kappa \sqrt{D_L D_0}$, for passive
pointlike particles \cite{Rosen}. Such a behavior persists for $v_0$ up to an
upper value, which appears to agree well with our estimate for $v_c$ in Eq.
(\ref{vc}). This picture holds also at lower thermal noise strengths,  Fig.~\ref{F3}(b) and Appendix C (though not with as good statistics). This
result confirms that for $v_0<v_c$ the array's edges make the JP
self-propulsion velocity, $\vec{v}_0$, to line up with the advection drag,
$\vec{v}_\psi$, so that the JP diffuses only through the FBL network due to
thermal fluctuations, .
% We remind that the $D_0$ dependence of $v_c$ in Eq.~(\ref{vc}) results from
%the estimated FBL thickness, $\delta=(D_0/\Omega_L)^{1/2}$. According to our
%derivation of $v_c$, we expect that the p(x) peaks for $v_0<v_c$ must shrink
%upon decreasing $D_0$. This is indeed the case, as shown in SI.

The intermediate regime, $v_c < v_0 \lesssim U_0$, is characterized by a
sharp drop of the particle's diffusivity. This is a signature of its pinning
to the array's edges. For $D_\theta/\Omega_L \ll v_0/U_0$, the particle can
slide along the edges only by overcoming the advection washboard potential of
amplitude $D_L|\langle \cos(2\pi y/L)\rangle|$. In the limit of very low
noises, $D_0/D_L, D_\theta/D_L \to 0$, this occurs for $v_0\sim U_0$. For
$D_0/D_L \ll \{|\langle \cos(2\pi y/L)\rangle|, v_0/U_0\}$, its diffusion
constant drops to exponentially small values \cite{Risken}, which could not
be computed numerically. On increasing $D_0$ and (or) $D_\theta$, the
amplitude of the pinning potential diminishes, and the particle's diffusivity
becomes numerically appreciable; eventually, the diffusion dips because
pinning becomes negligible.

Interesting is the shift of the $D$ minima to
higher $v_0$ values with increasing $D_0$ (inset of Fig.~\ref{F3}). This
counterintuitive effect, is due to the fact that for $D_\theta/\Omega_L <
D_0/D_L \ll 1$, the JP self-propulsion velocity, $\vec{v}_0$, changes
direction owing to the combined action of thermal noise (pulling the particle
away from its pinning site) and advection (exerting a torque on it). A larger
dispersion of the JP orientation angle, $\theta$, with thermal noise, implies
a higher depinning value of $v_0$.

\begin{figure}[tp]
\centering \includegraphics[width=8cm]{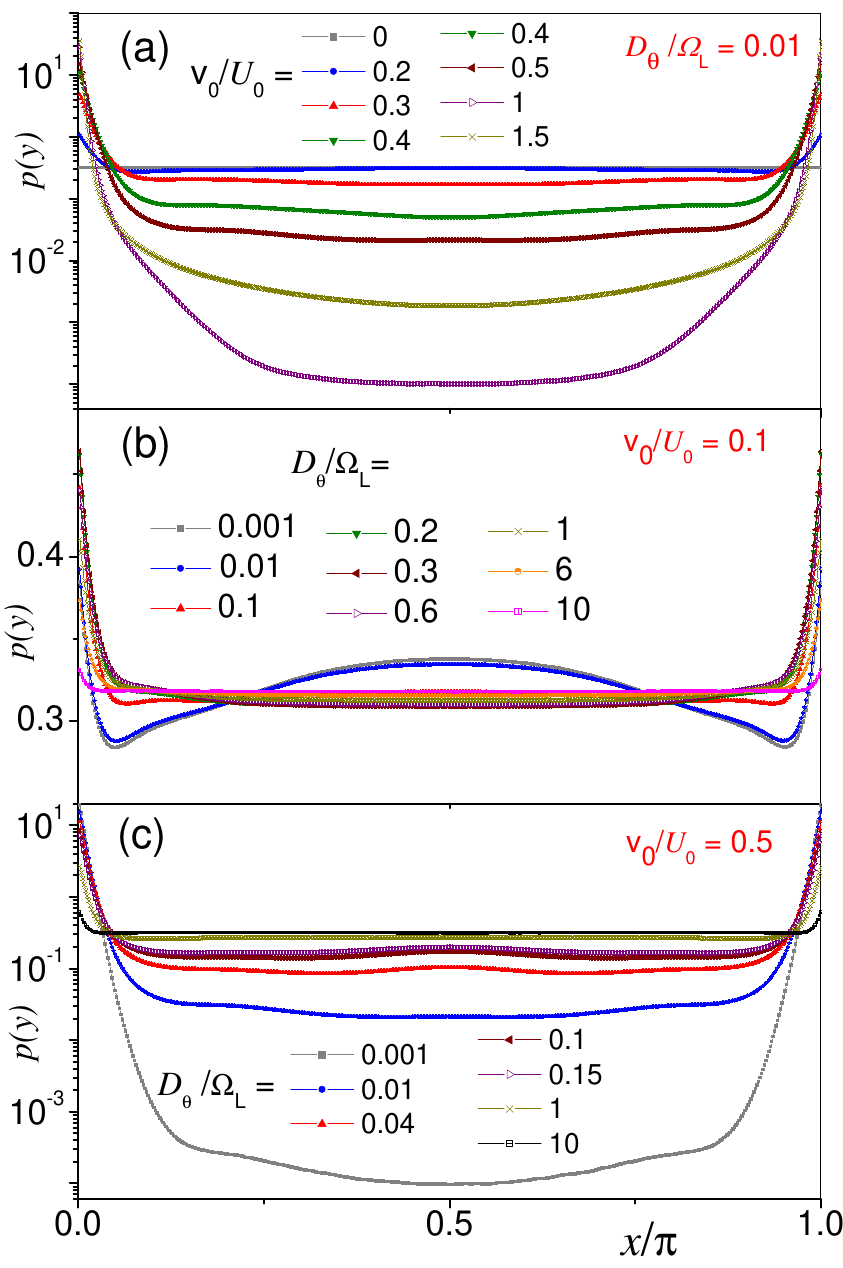}
\centering \includegraphics[width=8cm]{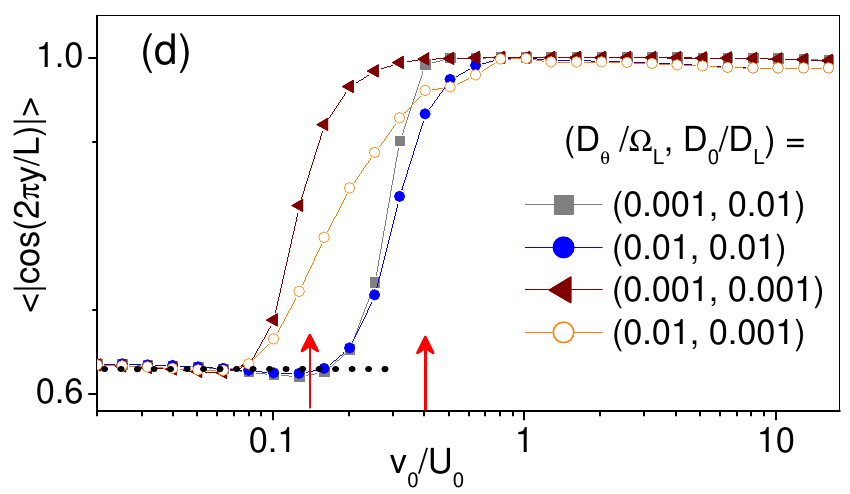}
\caption{Stationary transverse distributions: (a)-(c) transverse density functions,
$p(y)$, corresponding to the
longitudinal distributions, $p(x)$, of Fig.~\ref{F2}; (d) $\langle |\cos (2\pi y/L)|\rangle$ vs.
$v_0/U_0$. Simulation parameters in (a)-(c)
are the same as in the corresponding panels of Fig.~\ref{F2}; if not specified otherwise in the legend,
the same parameters have been adopted in (d). Vertical
arrows mark our estimates for
$v_c$ at $D_0=0.01$ and $0.001$, Eq.~(\ref{vc}).}
\label{F5}
\end{figure}

We stress here, once again, the role of advection along the array's edges. In
a periodic 2D convection array of stream function (\ref{psif}), the
$v_0$-dependence of $D$ is quite different \cite{Neufeld,RR2}. In the
noiseless limit, a spherical JP gets trapped for $v_0$ lower than the
threshold $v_{\rm th}\simeq 2.2~ U_0$ \cite{Neufeld}. After a generally long
transient, during which it keeps roll jumping, the particle eventually ends
being uniformly distributed inside a single convection roll (i.e, with
$v_c=0$). Here, instead, advection at the array's edges lowers the trapping
threshold down to $U_0$.

As anticipated above and illustrated in Fig.~\ref{F4}, 
$D_\theta$, is an important control parameter, because it governs orientation
and persistence of self-propulsion. When $D_\theta$ is so large that $D_s$ is
negligible compared to $D_0$, $D_\theta/\Omega_L \gg (D_L/D_0)(v_0/U_0)^2$,
the passive particle regime is recovered, no matter the value of $v_0$. In Fig.~
\ref{F4} we set $D_0 < D_L$, i.e., ${\rm Pe} \gg1$: therefore, for $D_\theta
/\Omega_L \to \infty$, all $D$ curves plotted there tend to the same (high
P\'eclet number) advection enhanced diffusivity value, $D=\kappa \sqrt{D_L D_0}$. More remarkably, the
two curves with $v_0<v_c$ only slightly deviate from that value throughout
the entire $D_\theta$ domain. This result is a further evidence of the
particle's confined circulation inside the FBLs.

The curves for $v_c < v_0 \lesssim U_0$ overshoot the advection enhanced diffusivity value, with
overlapping maxima at $D_\theta \sim \Omega_L$. This effect is due to the
synchronized action of angular diffusion and advection torque. The two,
combined, optimize the mechanism of edge switching, whereby the JP moves from
one pinning site at the bottom to a pinning site at the top, and vice versa.
As such pinning sites are (at least) a distance $L/2$ apart, edge switching
enhances lateral diffusion against edge pinning. Under these conditions (Fig.~
\ref{F4}), the leading contribution to the diffusion constant is
$D=(L/2\pi)^2 \Omega_L/2$, i.e., $D/D_L \simeq 1/2$, independent of $v_0$ and
$D_0$ (Appendix A).

The curves for $v_0 \gg U_0$, as anticipated above, are mostly governed by
self-propulsion. For a wide $D_\theta$ range, they closely follow the free
diffusion law, $D=D_0+D_s$, like in a straight zero-flow channel, even when
$l_\theta>L/2$ (as a consequence of the reflecting boundaries). However,
having chosen $D_0\ll D_L$, at larger $D_\theta$ the JP free diffusion
constant, $D_0+D_s$, grows smaller than $D_L$: Particle's diffusion then
takes place at effective high P\'eclet numbers and the advection enhanced 
diffusivity mechanism applies,
hence $D=\kappa\sqrt{D_L(D_0+D_s)}$ \cite{RR2}. These two distinct diffusion
laws are both illustrated in Figs. \ref{F3} and \ref{F4}.

All $D$ curves with $v_0>v_c$ in Fig.~\ref{F4} share a surprising property:
Upon lowering $D_\theta$, they drop below the free diffusion value, $D_0$.
This suggests that for $v_0>v_c$ and large $\tau_\theta$ advection at the
array edges is never negligible. We already noticed that a JP with
$l_\theta>L/2$ trapped at the array's edges can free itself either by sliding
against the advection drag or by switching edge. We also know that in the
limit $D_0/D_L, D_\theta/D_L \to 0$, a particle pointing against an edge with
$|\cos \theta|< U_0/v_0$ ends up sitting in a stagnation corner, i.e.,
sliding can be suppressed even for $v_0 \gg U_0$. Diffusion is then activated
by autonomous edge switching, which, for a JP, can occur through either
angular reorientation, with time constant $\tau_\theta$, or translational
diffusion, with time constant $\tau_D=(L/2\pi)^2/2D_0$ (Appendix A). For $v_0
\gg U_0$, depinning from the edge washboard potential requires $|\theta|>
U_0/v_0$, therefore, angular diffusion ceases driving edge switching when
$2D_\theta \tau_D < (U_0/v_0)^2$, or $D_\theta < D_\theta^*$, with
\begin{equation} \label{Db}
D_\theta^*/\Omega_L =(U_0/v_0)^2 (D_0/D_L).
\end{equation}
As shown in Fig.~\ref{F4}, lowering $D_\theta$ below $D_\theta^*$ causes a
sharp drop of the $D$ curves to values so small that they could not be
numerically computed with acceptable accuracy. This effect is clearly due to
geometric confinement, as confirmed by the fact that it was never detected in
2D flows \cite{RR2}.

\section{Concluding remarks} \label{Conclusions}

We have investigated the diffusion of an active JP in a 1D array of
counter-rotating convection rolls. The JP considered here should be regarded
as modeling a self-propelling micro-swimmer of biological or synthetic
nature, alike. Our choice for the laminar flow is meant to mimic the
Rayleigh-B\'enard rolls occurring between two parallel surfaces kept at an
appropriate temperature difference.

We focused on effects due the combination of three key ingredients, namely,
thermal noise, advection and self-propulsion, in a confined geometry. Such
effects, not detectable in 2D arrays of convection rolls with same
hydrodynamical parameters but no boundaries, can be summarized as follows:
\newline(i) The large-scale circulation of a JP trapped in a convection roll is
confined to narrow flow boundary layers, whereby the particle self-propulsion
velocity tends to line up with the advection drag, which results in an
accumulation of the particle probability density. \newline(ii) The diffusion of an
active JP with low self-propulsion speed is governed by its circulation along
the roll boundaries, and is thus undistinguishable from that of a regular
passive particle. \newline (iii) For larger self-propulsion speeds, the JP tends to
sojourn in the vicinity of the array's edges and diffuses by sliding along
them. Its diffusion is dominated by the advection drag parallel to the
array's boundaries even for self-propulsion speeds much larger than the
advection drag. This mechanism works for strengths of the angular noise above
a certain threshold; below that threshold, the particle's diffusion constant
drops to vanishingly small values.

Our emphasis on the above confinement effects is motivated by the widespread
interest in controlling transport of diluted active matter
\cite{Muller,Marchetti} in microfluidic circuits \cite{Kirby}.

To this regard we note that,  due to the large variability of the advection
parameters in actual Rayleigh-B\'enard cells \cite{Tabeling,Gollub1} [$L$ and $U_0$ in the model
of Eq. (\ref{psif})] and the self-propulsion mechanisms \cite{Gompper,Wang} [$v_0$ and
$D_\theta$ in the JP model of Sec. \ref{Model}], all three diffusion regimes
listed above are experimentally accessible.
Both organic and synthetic microswimmers could be employed to investigate
advection effects on the active diffusion in laminar flow patterns. 

Finally, we remark that the overall picture presented here holds for rigid (or no-slip)
boundary arrays, as well. Numerical results for the stream function,
\begin{equation}
\label{psir}
\psi(x,y)= ({U_0L}/{2\pi})\sin({2\pi x}/{L})\sin^2({2\pi y}/{L}),
\end{equation}
are reported in Figs. {\ref{F3} and \ref{F4} for a comparison. The breakdown
of the FBL circulation with increasing $v_0$ is still detectable, though not
as sharp as in free-boundary convection arrays. In Fig.~\ref{F3} the dips of
the $D$ curves occur at lower values of $v_0$ and are less pronounced. This
happens because a particle moving against the edges of the array of Eq.
(\ref{psir}) is advection free: it moves subjected to the sole thermal noise;
correspondingly, the FBL width shrinks. For the same reason, in Fig.~\ref{F4}
the $D$ curves never drop below $D_0$.

\appendix

\section{Model's time scales}\label{A}

The diffusion process of Eq.~(\ref{LE}) is characterized by many dynamical
parameters. In particular, in our analysis of the diffusion data we made use
of various time scales, which we now recap for reader's convenience, with
reference to the underlying dynamical mechanisms:

{\it (i) Angular diffusion.} In Eq.~(\ref{LE}), the self-propulsion velocity
vector, ${\vec v}_0$, was assumed to have constant modulus, $v_0$, and
fluctuating orientation with angle $\theta(t)$. In the absence of advection,
$U_0=0$, we know that \cite{ourPRL,Golestanian}, $\langle v_i(t)
v_i(0)\rangle =(v_0^2/2)\exp(-D_\theta |t|)$, with $i=x,y$. These
autocorrelation functions prove that the angular noise strength, $D_\theta$,
plays the role of angular diffusion rate and, accordingly,
\begin{equation}
\label{A1}
\tau_\theta=1/D_\theta,
\end{equation}
defines the persistence time of the ensuing active Brownian motion of the
self-propelling JP.

\begin{figure}[tp]
\centering \includegraphics[width=6.7cm]{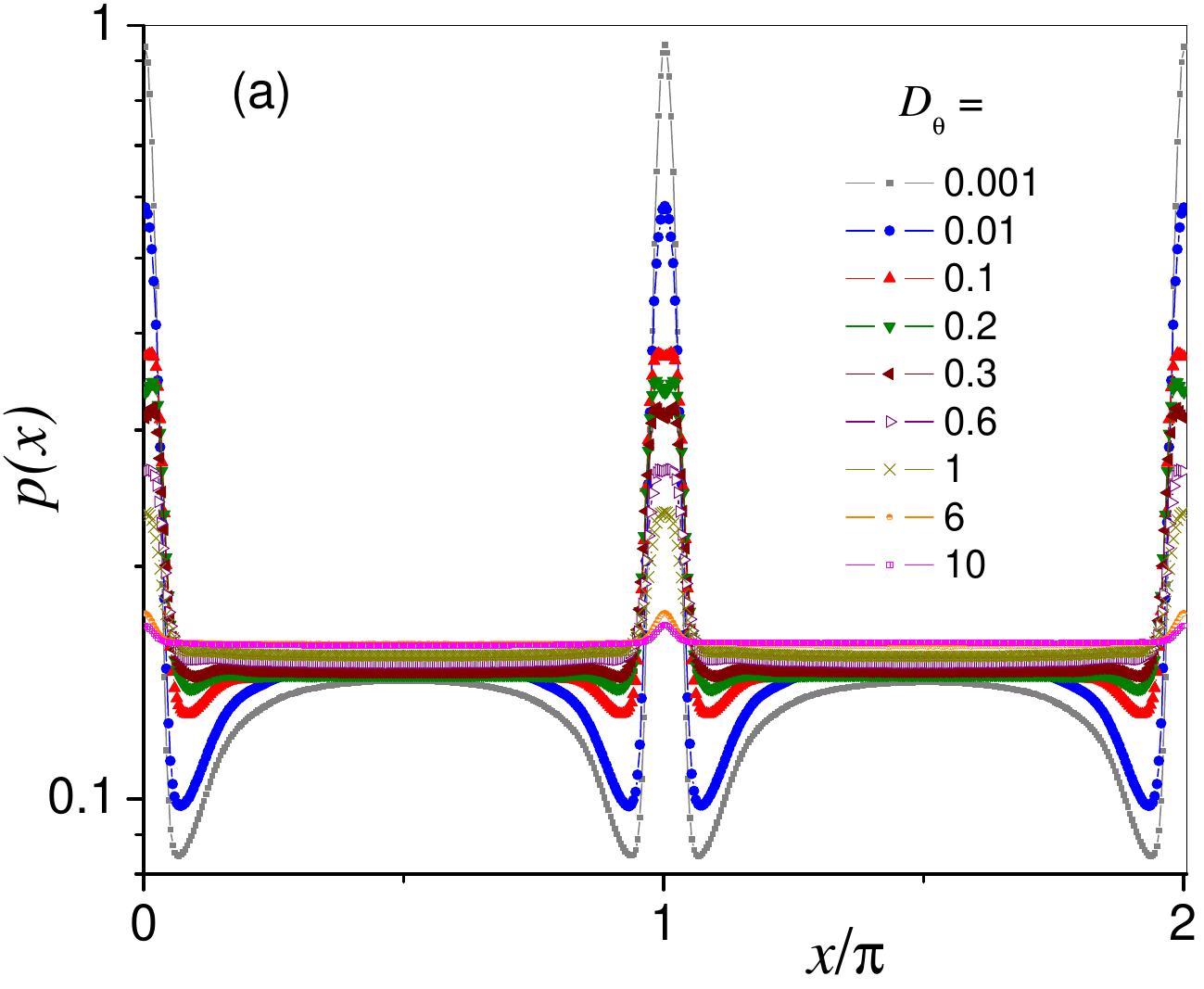}
\centering \includegraphics[width=7cm]{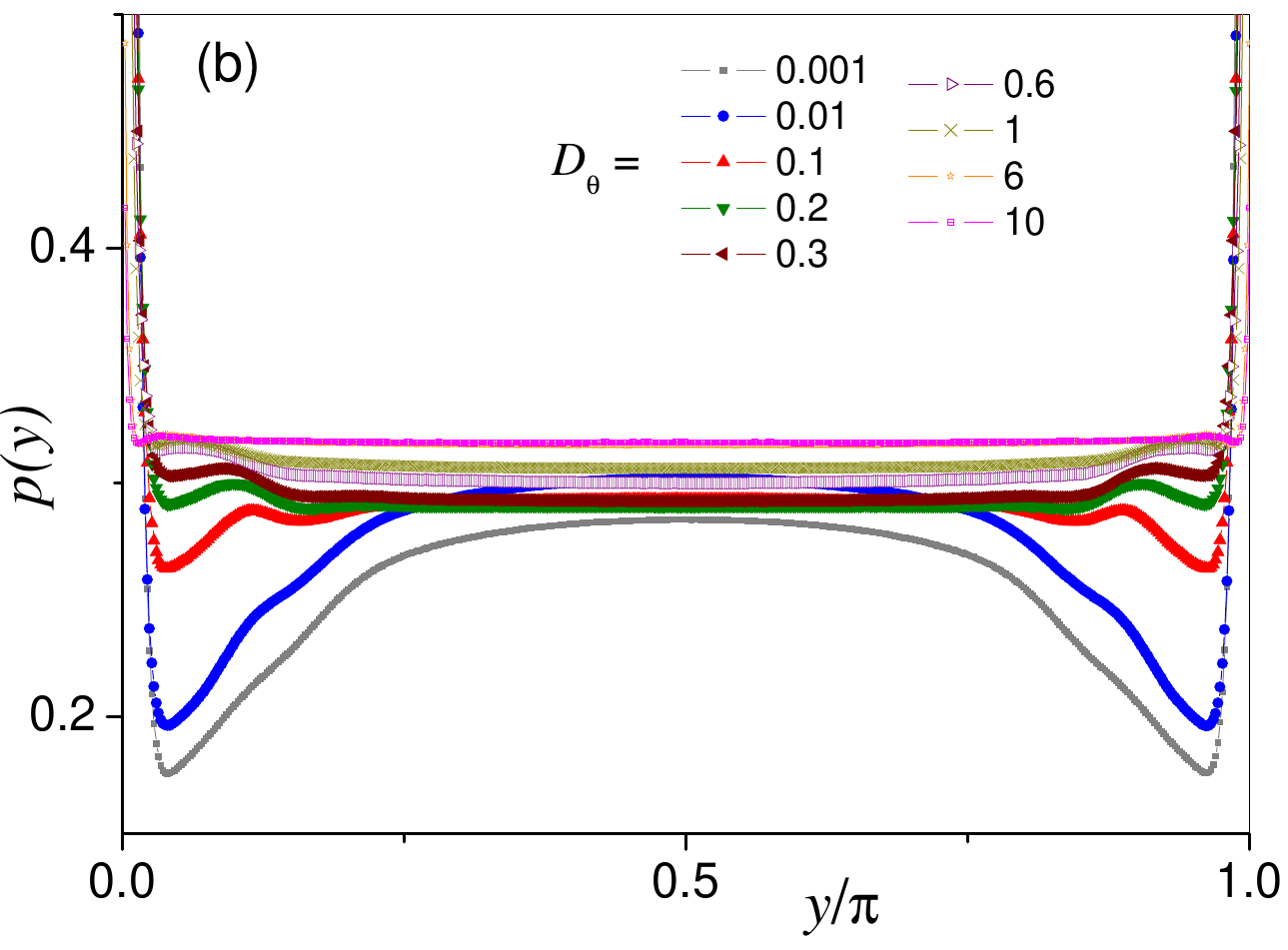}
\caption{Stationary distributions, $p(x)$ and $p(y)$, for different $D_\theta$
with the same simulation parameters as in
panels (b) of Figs. \ref{F2} and \ref{F5}, except for $D_0=0.001$. Note that here
Eq.~(\ref{vc}) yields $v_c = 0.13$.}
\label{F6}
\end{figure}

{\it (ii) Roll circulation.} Due to advection, a particle trapped in a
convection roll, is dragged along a circular FBL of approximate radius $L/4$
with speed $U_0$. This means that the particle circulates inside the trapping
roll with period of the order of $\tau '_L =\pi L/2U_0$ or, equivalently,
angular frequency $\Omega '_L=4U_0/L=(2/\pi)\Omega_L$. Therefore,
consistently with the current literature, we agreed to use the standard
definition of circulation time scale \cite{ourPOF}, namely
\begin{equation}
\label{A2}
\tau_L=2 \pi/\Omega_L =L/U_0.
\end{equation}

{\it (iii) Thermal diffusion.} Subjected to thermal noise, the suspended
particle diffuses across the array with mean first-passage time \cite{Redner}
$\tau '_D=(L/2)^2/2D_0$. Advection drag and thermal diffusion are comparable
when $\tau '_D/\tau '_L\sim 1$. In the text, this condition has been
reformulated more conveniently as $\Omega_L \tau_D \sim 1$, with
\begin{equation}
\label{A3}
\tau_D=(L/2\pi)^2/2D_0.
\end{equation}

{\it (iv) Ballistic self-propulsion.} In the absence of angular diffusion,
$D_\theta=0$, the JP crosses ballistically a unit flow cell with time
constant $\tau '_0=L/\langle|v_x|\rangle=\pi L/2v_0$. In this regime, the
action of advection and self-propulsion are comparable under the condition
that $\tau '_0 \sim \tau '_L$, or, equivalently,  $\tau_0 \sim \tau_L$, with
\begin{equation}
\label{A4}
\tau_0=L/v_0.
\end{equation}
It should be noted that ballistic effects due to self-propulsion are
negligible with respect to advection and the array's geometry, respectively
under the conditions $\Omega_L \tau_0 \gg 1$ and $D_\theta \tau_0 \gg 1$,
that is, for $v_0\ll U_0$ and $D_\theta/\Omega_L \gg v_0/U_0$ \cite{RR2}.

Equations (\ref{A1})-(\ref{A4}) define the time scales used in our analysis
of the simulation data displayed in Figs. \ref{F3} and \ref{F4}. They can
also be combined to obtain convenient estimates of the reference diffusion
scales introduced in Sec. \ref{Dx}. Firstly, based on our derivation of
$\Omega '_L$, the diffusing particle is advected across the array width $L/2$
with effective speed $(2/\pi)U_0$ \cite{CPL}. This means that by hitting the
roll boundaries it undergoes a large-scale diffusion with diffusion constant
$D=(1/2)(L/2)(2U_0/\pi)=D_L$, which coincides with the diffusion scale
associated with the stream function  of  Eq.~(\ref{psif}). Secondly, in Sec.
\ref{px}, the FBL of a convection roll has been modeled as an annulus of
radius $L/4$ and width $\delta=(D_0/\Omega_L)^{1/2}$; accordingly, it covers
a fraction $\phi=2\pi \delta /L$ of the roll's surface. We know that, for
$v_0<v_c$, the large-scale diffusion of a JP, with diffusion constant $D_L$,
is restricted to the FBL network. Therefore, its effective diffusion constant
is $D=\phi D_L$, that is, $D=\sqrt{D_L D_0}$. This simple argument reproduces
the result of Ref. \cite{Rosen} with $\kappa=1$ instead of the more accurate
$\kappa=1.07$.

\begin{figure}[tp]
\centering \includegraphics[width=7.0cm]{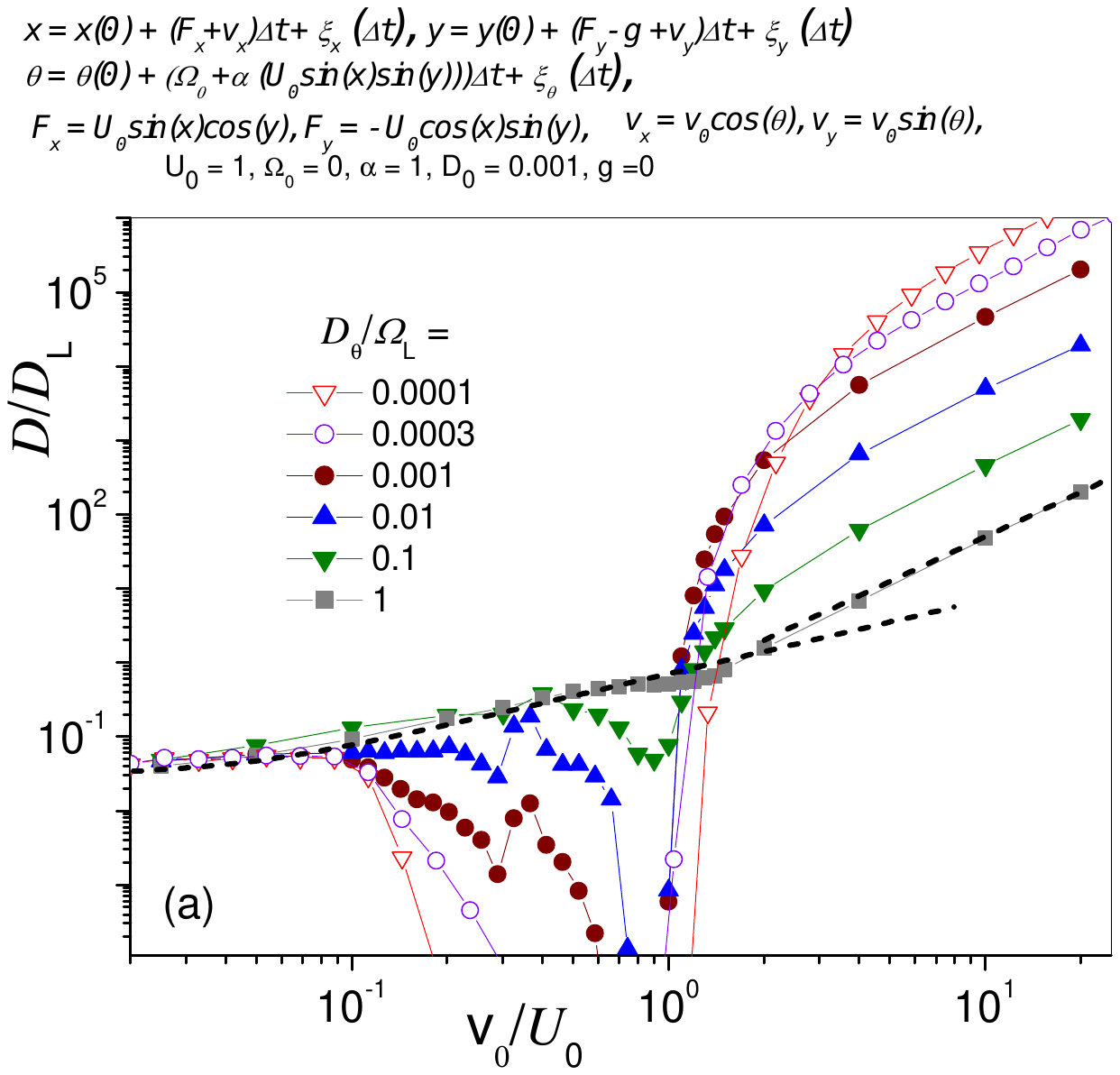}
\centering \includegraphics[width=7.0cm]{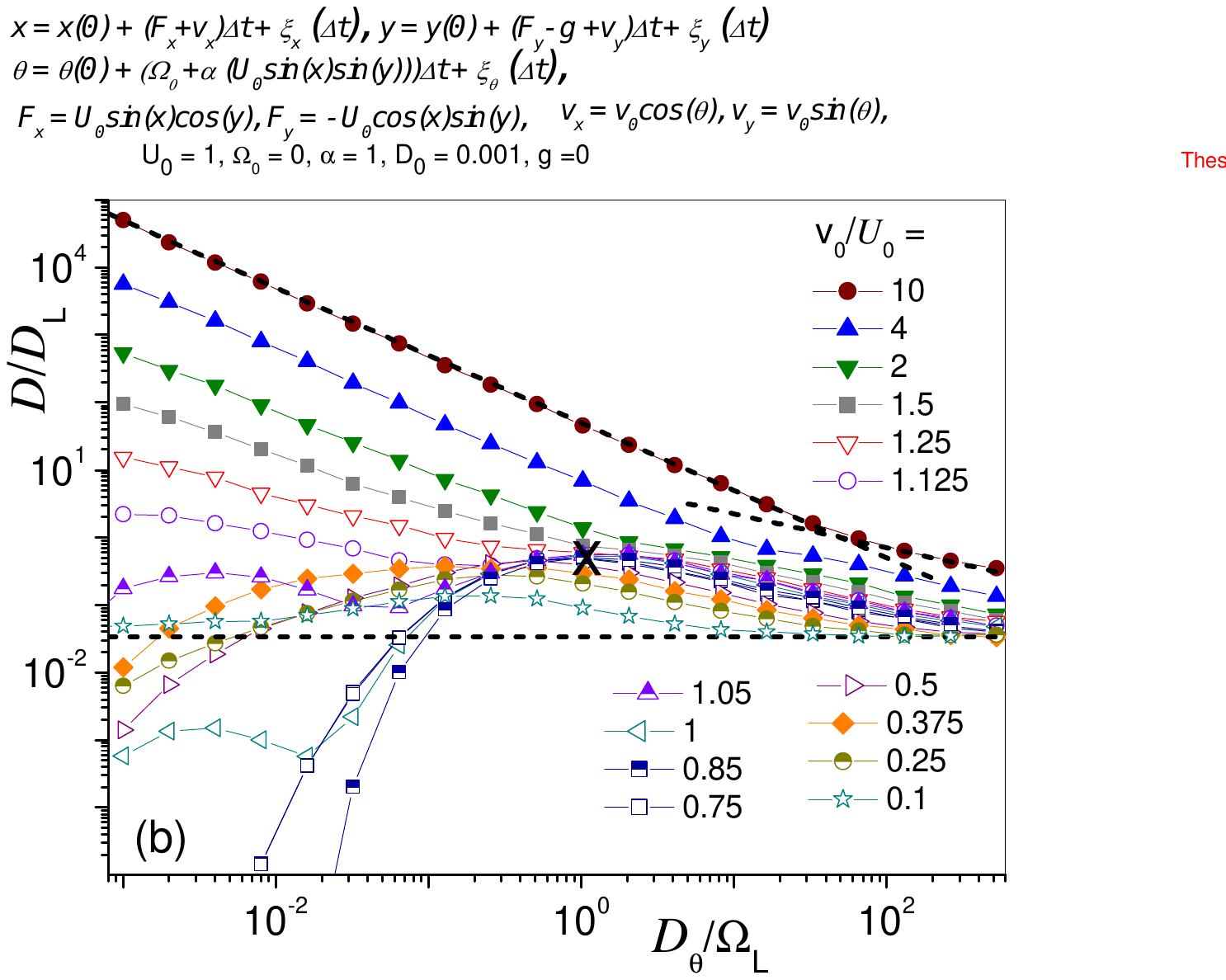}
\caption{Longitudinal diffusion of a JP in the laminar flow of Eq.~(\ref{psif}):
same as in Fig.~\ref{F3}, (a), and  Fig.~\ref{F4}, (b),
but for $D_0=0.001$.} \label{F7}
\end{figure}

\section{Transverse distributions}

We present in Fig.~\ref{F5}  the transverse distributions, $p(y)$,
corresponding to the longitudinal distributions, $p(x)$. Combined with Figs.
\ref{F1}(c),(d) and \ref{F2} of Sec. \ref{px}, this figure illustrates the
large-scale circulation of a JP with $v_0<v_c$ and the break-up of the FBLs
for $v_0>v_c$. The depletion of the inner region of the convection rolls is
the most pronounced for $v_0 \simeq U_0$ [Fig.~\ref{F5}(a)], which
corresponds to the strongest interplay of advection and self-propulsion. In
Fig.~\ref{F5}(d), the FBL break-up causes the sharp jumps of $\langle |\cos
(2\pi y/L)|\rangle$, from $0$ to $1$ at $v_0\sim v_c$.

Moreover, we stated in Sec. \ref{px} that the flattening of the $p(x)$ for
$v_0 \gg U_0$ is due to the symmetric particle accumulation against both
array's edges. That statement is supported here by the profile of the
corresponding $p(y)$ curves of Fig.~\ref{F5}(a), which, indeed, exhibit sharp
maxima at $y=0$ and $y=L/2$.

We remind once again that the nonuniform distributions $p(x)$ and $p(y)$ are
peculiar of 1D convection arrays. Indeed, particle accumulation inside the
FBLs for $v_0<v_c$ and against the array's edges for $v_0>v_c$ is an effect
of geometric confinement. Numerical simulations of an active JP in the 2D
flow of Eq.~(\ref{psif}), with periodic boundary conditions in the $x$ and
$y$ direction, returned uniform longitudinal and transverse distribution for
any value of $v_0$ (not shown).

\section{The role of thermal noise}

For brevity, in Secs.~\ref{px} and \ref{Dx} we did not dwell on the role
thermal noise. We just stressed that its strength, $D_0$, was set much
smaller than the advection diffusion scale, $D_L$. Accordingly, we defined
the P\'eclet number as ${\rm Pe} = D_L/D_0$. We then mentioned that $D_0$
enters our estimates of both the FBL width, $\delta = (D_0/\Omega_L)^{1/2}$,
and the break-up threshold, $v_c$, in Eq.~(\ref{vc}).

To support those statements we present here simulation results for the 1D
distributions, $p(x)$ and $p(y)$, and the asymptotic diffusion constant, $D$,
obtained for a value of $D_0$ one order of magnitude smaller than in Figs.
\ref{F2}-\ref{F4}. On comparing Fig.~\ref{F6}(a) with Fig.~\ref{F2}(b), it is
apparent that the FBL width shrinks with increasing $D_0$. Analogously, the
existence of the threshold $v_c$ and its dependence on $D_0$ are confirmed by
the curves $D$ versus $v_0$, in Fig.~\ref{F7}(a), and $D$ versus $D_\theta$,
in Fig.~\ref{F7}(b) [see also Fig.~\ref{F5}(d)].

The overall behavior of the diffusion curves in Figs. \ref{F7} is consistent
with that displayed in Figs. \ref{F3} and \ref{F4}. For instance, in Fig.~
\ref{F7}(b) all curves with $v_c < v_0 \lesssim U_0$ attain the same maximum,
$D/D_L\simeq 1/2$ at $D_\theta/\Omega_L=1$, as in Fig.~\ref{F4}, i.e.,
independently of $D_0$. However, a few differences are worthy to note: (i)
Diffusion in the pinning range, $v_c < v_0 \lesssim U_0$, of Fig.~\ref{F7}(a)
reveals additional details, which went unnoticed in Fig.~\ref{F3}; (ii) These
details reflect into the non-monotonic $D_\theta$-dependence of the
corresponding $D$ curves in Fig.~{\ref{F7}}(b); (iii) The interplay between
thermal, $D_0$, and angular noise, $D_\theta$, causes the double-peaked
aspect of the $p(x)$ maxima in Fig.~\ref{F6}(a) [absent in Fig.~\ref{F2}(b)].
These details do not affect the main conclusions of our work. We also remark
here that obtaining simulation data with a good statistics at very low noise
levels, $D_0 \to 0$ and (or) $D_\theta \to 0$, would require exceeding
computational resources. For this reason we could not push our numerical
investigation to lower $D_0$ values.

A more substantial difference between Figs.~\ref{F4} and \ref{F7}(b) regards
the curves with $v_0 \gg U_0$. In Fig.~\ref{F7}(b) those curves seem to not bend
downward upon decreasing $D_\theta$. This is due to the fact that,
consistently with Eq.~(\ref{Db}), for the simulation parameters of Fig.~\ref{F7} the estimated position of their maxima, $D^*_\theta$, is not
captured by the numerically accessible $D_\theta$ range.

\section*{Acknowledgements}
We thank RIKEN Hokusai for providing computational resources. 
Y.L. is supported by the NSF China under grants No. 11875201 and No.
11935010. P.K.G. is supported by SERB Start-up Research Grant (Young
Scientist) No. YSS/2014/000853 and the UGC-BSR Start-Up Grant No.
F.30-92/2015. F.N. is supported in part by: NTT Research,
Army Research Office (ARO) (Grant No. W911NF-18-1-0358),
Japan Science and Technology Agency (JST)
(via the CREST Grant No. JPMJCR1676),Japan Society for the Promotion of Science (JSPS) 
(via the KAKENHI Grant No. JP20H00134 and the JSPS-RFBR Grant No. JPJSBP120194828),
the Asian Office of Aerospace Research and Development (AOARD) (via Grant No. FA2386-20-1-4069),
and the Foundational Questions Institute Fund (FQXi) via Grant No. FQXi-IAF19-06.

\end{document}